\documentclass[12pt]{iopart}
\usepackage{amsbsy,amsfonts,amscd,amstext,suppl}

\newfont{\eu}{eusb10 at 12pt}
\newcommand{\Rset}{\mathbb{R}}
\newcommand{\sgt}{\tilde{\sigma}}
\newcommand{\bfcM}{{\boldsymbol{C}_0^1}}
\newcommand{\bfcm}{{\boldsymbol{c}_0^1}}
\newcommand{\hf}{{\hat f}}

\newcommand{\tom}{T_{p_0}\m}
\newcommand{\omi}[2]{\omega_{#1#2}}
\newcommand{\oms}[2]{\omega^{#1#2}}

\newcommand{\gam}[3]{\gamma^#1_{#2#3}}
\newcommand{\diff}[1]{Dif\!f{}_{\mathrm{loc.}}^{\mathrm{#1}}}
\newcommand{\sg}{{\sigma}}
\newcommand{\GammaP}{\Gamma_{\mathrm{G}}}
\newcommand{\GammaG}{\Gamma_{\mathrm{\widehat{G}}}}
\newcommand{\SGammaH}{\Gamma_{\mathrm{\widehat{H}}}}

\newcommand{\m}{{\mathcal{M}}}
\newcommand{\Sub}{{\mathcal{S}}}
\newcommand{\Su}{{\mathcal{S}^1_c}}
\newcommand{\A}{{\mathcal{A}}}
\newcommand{\B}{{\mathcal{B}}}
\newcommand{\F}{{\mathcal{F}}}
\newcommand{\G}{{\mathcal{G}}}
\newcommand{\I}{{\mathcal{I}}}
\newcommand{\J}{{\mathcal{J}}}
\newcommand{\Cc}{{\mathcal{C}}}

\newcommand{\Lag}{{\mathcal{L}}}
\newcommand{\Sz}{{\mathcal{S}^0_c}}
\newcommand{\tsms}{\underline{T^\ast\Rset^n}}
\newcommand{\Oc}{\widehat{\Omega}}

\newcommand{\kc}{\hat{\kappa}}
\newcommand{\bfc}{{\boldsymbol{c}}}
\newcommand{\bfah}{{\boldsymbol{\hat{a}}}}
\newcommand{\bfa}{{\boldsymbol{a}}}

\newcommand{\bftau}{\boldsymbol{\tau}}
\newtheorem{theorem}{Theorem}

\newtheorem{definition}{Definition}

\begin{document}

\title[Spacetime deployment]{Spacetime deployments 
para\-metri\-zed by gravitational and electromagnetic fields}

\author{Jacques L. Rubin\footnote{jacques.rubin@inln.cnrs.fr}
and Thierry Grandou\footnote{thierry.grandou@inln.cnrs.fr}}

\address{Institut du Non-Lin{\'e}aire de Nice, U.M.R. 6618
C.N.R.S. - Universit{\'e} de Nice - Sophia-Antipolis,  1361
route des Lucioles, 06560 Valbonne, France.}

\begin{abstract}
We consider
spacetimes with measurements of conformally invariant physical
properties. Then, applying the Pfaff theory for PDE to a
particular conformally equivariant system of differential equations,
we make explicit the dependence of any kind of function describing  a
``spacetime deployment", on $n(n+1)$ parametrizing functions,
denoting by $n$ the spacetime dimension. These functions,
appearing in a linear differential Spencer sequence can be
consistently ascribed to unified electromagnetic  and
gravitational fields, at any spacetime dimensions
$n\geq4$.\par\bigskip
\end{abstract}

\maketitle
%%%%%%%%%%%%%%%%%%%%%%%%%%%%%%%%%%%%%%%%%%%%%%%%%%%%%%
%%%%%%%%%%%%%%%%%%%%%%%%%%%%%%%%%%%%%%%%%%%%%%%%%%%%%%
%%%%%%%%%%%%%%%%%%%%%%%%%%%%%%%%%%%%%%%%%%%%%%%%%%%%%%
\section{Introduction: spacetime and navigation}
\subsection{Substrat and unfolded spacetimes} 
In the present article, most part of the matter is based on a
particular type of gravity known under the spell of
``conformal gravity". Actually we will adopt throughout a
specific Equivalence Principle which has recently been
formulated by M. Ghins and T.~Budden \cite{ghins}, and from
which we think that the very conformal aspect of our
approach proceeds. This so-called Punctual Equivalence
Principle (PEP) can be stated as follows \cite[p. 44
but with the notation $p_0$ instead of $p$]{ghins}:
\begin{quote}
\textbf{Punctual Equivalence Principle (PEP):}
for all $p_0\in\m$, special relativity holds at
$p_0$  ``in the restricted sense".
\end{quote}
This is a local definition on a spacetime
manifold $\m$, and the ``in the restricted
sense"  should be given the meaning \cite[p. 43]{ghins}:
\begin{quote}
\textbf{Special relativity holds at $p_0$:} there exists a
local chart $x^\mu$ of a neighbourhood of $p_0$ such that the
fundamental dynamical and curvature-free special relativistic
laws hold in their standard vectorial form in $x^\mu$ at
$p_0$.
\end{quote}
The coordinate maps $x^\mu$ ($\mu=1,\dots,\dim\m$) are local
charts defined on a neighbourhood $U(p_0)$ of a given point
$p_0$ in $\m$.\par
Indeed, a peculiar type of charts is selected, made out of
those charts which, to a given $U(p_0)$, associate a
neighbourhood of the origin of the vectorial space, $\tom$, tangent  to
$\m$ at $p_0$. This local, punctual approach can
certainly be intuitively motivated different ways. Let us
just mention that it seems intimately related to some of the
practical aspects met in the situations of satellites
navigation such as the GPS or GALILEO systems. In
effect, each of the satellites belonging to a GPS
constellation, so to say, realizes somehow this kind of a
local punctual equivalence by registering other satellites own
ephemeris and calandar data, as well as proper times given by
embarked atomic clocks. In this respect, one can say that
spatio-temporal charts are achieved ``on board", on some kind
of ``table of the charts" $\tom$, as one would say with the
help of a navigation terminology broadly used in the GPS
technologies. From the mathematical standpoint, it will
therefore be postulated that these ``on board" or ``table of
the charts" aspects, are those of a running vectorial
tangent spacetime $\tom$.\par
One has indeed to realize a set of charts of the
projective space defined on, and associated to each moving
tangent spacetime. That is, in particular, and at a given
fixed proper time, charts of the celestial sphere attached to
a given observer at $p_0$. This, in turn, amounts to recognize
that the charts are implicitly associated to a conformal
geometry.\par
In order to stress the fundamental deployment aspect of our
approach, the following terminology will be used throughout:
\begin{itemize}
\item $\tom$, the  spacetime tangent to $\m$ at $p_0$, will
be referred to as the underlying  or substrat
spacetime $\Sub$.
\item From this substrat spacetime, the unfolded spacetime
$\m$ will be envisaged and referred, by means of the Punctual
Equivalence Principle at $p_0\in\m$.
\end{itemize}
It may be worth stressing that these moving tangent
spacetimes, $\tom$, are those spaces in which the conformal
physical measures will be achieved some way or other, with
the help of rods, compass, clocks, recorders, etc\dots
Eventually, all of the foregoing considerations may be
adapted to a variety of pretty different situations as the
most essential aspect wil be the one of a spacetime manifold
deployment ($\m$) from a substrat spacetime ($\Sub$). We build up
the mathematical formalism corresponding to the deployment of
a conformal geometric structure from an isometrical  one.

%%%%%%%%%%%%%%%%%%%%%%%%%%%%%%%%%%%%%%%%%%%%%%%%%%%%%%
%%%%%%%%%%%%%%%%%%%%%%%%%%%%%%%%%%%%%%%%%%%%%%%%%%%%%%
%%%%%%%%%%%%%%%%%%%%%%%%%%%%%%%%%%%%%%%%%%%%%%%%%%%%%%

\subsection{To tie a spacetime ship with its
environnement: the principle of equivalence}
We assume the unfolded spacetime $\m$  to be of class
$C^\infty$, of dimension $n\geq 4$ and locally connected. Let $p_0$
be a particular point in
$\m$, $U(p_0)$ an open neighborhood of $p_0$ in $\m$, and $\tom$ its
tangent space. The so-called \emph{``punctual"\/} 
principle of equivalence we will be  relying on (compatible with the usual ones
from M. Ghins view; Private communication), states that it exists a local
diffeomorphism, 
$\varphi_{p_0}$,  we call the ``equivalence map", attached to
$p_0$, and putting in a one-to-one correspondence the points
$p\in U(p_0)$ with some vectors $\xi\in\tom$ in an open neighborhood of
the origin of $\tom$:
\[
\varphi_{p_0}:p\in
U(p_0)\subset\m\longrightarrow\xi\in\tom\,,\qquad
\varphi_{p_0}(p_0)=0\,.
\]
We will see that this description is really well-suited
since it is also strongly related to the mathematical tools to be used in
the sequel, where any given point $p_0$ acquires a
specific mathematical status.\par  
Let us add that a more standard local equivalence principle would
consist in considering
$\Sub$ an Euclidean space
$\Rset^n$ and the equivalence map $\varphi$ as a local chart of an
atlas of $\m$ on an open neighborhood $U$, such that:\par
\[
\varphi:p\in
U\subset\m\longrightarrow\xi\in\Rset^n\simeq\Sub\,.
\]
\par\smallskip
Moreover, we assume that each of these two kinds of spacetimes is
endowed with a metric field,  denoted by $g$ for $\m$,
and by $\omega$ with signature $(+,-,-,-)$ for $\Sub$.\par
Eventually, we make the general assumption that
$\Sub$ has a constant Riemaniann scalar curvature with value
$n(n-1)k_0$ ($k_0\in\Rset$), and then is ``conformally flat",
i.e., the  Weyl tensor is vanishing. This assumption, in
view of the H. Weyl theorem \cite{weylkonf}, will ensure the
integrability of the conformal Lie pseudogroup associated to
$\omega$, and denoted here by
$\GammaG$.\par Then if ${\hf}\in{\GammaG}$, ${\hf}$ is a solution  of the PDE
system:
\begin{equation}
\hf^\ast(\omega)=e^{2\alpha}\,\omega\,,
\label{first}
\end{equation}
with $\det(J({\hf}))\not=0$, 
$J({\hf})$  the Jacobian of $\hf$, and ${\hf}^{\ast}$ its
pull-back. Also, $\alpha$ is a function associated to, and varying with each
$\hf$. As a particular case, the set of diffeomorphisms $\hf$ for
which $\alpha=0$ constitutes the so-called ``Poincar{\'e} Lie
pseudogroup $\GammaP$" or, equivalently, the Lie pseudogroup of
isometries. We denote by
$f$ the elements of
$\GammaP$. Contrarily to appearances, an element $\hf$ can't
be defined in  a somewhat one-to-one correspondence, out of a
given element $f$ and a given function $\alpha$. This can
be obtained only if the metric field
$\omega$ on $\Sub$ satisfies a particular condition, that we will
call ``S-admissibility", which will be defined and precised
in the sequel. Moreover, if the latter condition is satisfied, then only
the elements of a proper Lie sub-pseudogroup of $\GammaG$, we
denote by $\SGammaH$, can admit such a decomposition, as will be
demonstrated at the exterior differential forms level (see
Theorem 3 below). This S-admissibility condition allows us to
define a deployment in the sense of a deformation from the
$\GammaP$ pseudogroup to the $\GammaG$ one.\par In fact,
since both spacetimes $\m$ and $\Sub$ are locally
diffeomorphic to
$\Rset^n$, the above alluded deployment relates  the
geometrical structures, i.e., their metric fields. Then we
consider that the metric field
$\nu\equiv\varphi_{p_0}^{-1\ast}(g)$ on $\tom$, is a
deployment or a deformation of
$\omega$. The  deployment or deformation terminology will
accordingly be used in either cases of $\m$ and $\Sub$, or
$\nu$ and $\omega$.\par We have to focus on the fact that
though  spacetimes $\m$ and $\Sub$ are diffeomorphic in view
of the equivalence map, their two respective metric fields 
are not, i.e., $g$ is not a pull-back of
$\omega$. Nevertheless these two metric fields will be conformally equivariant. 
In some analogy with the decomposition for $\hf$, the metric
field $\nu$ will be described in terms of $\omega$ and some
other fields which will reveal to be thinkable in terms of
unified electromagnetic and gravitational fields. In fact,
the deformations  in
$\SGammaH$ of the applications $f\in\GammaP$ will define tetrads used to
obtain $\nu$ out of $\omega$. Hence, as a result, the equivalence map will
be also parametrized by electromagnetic and gravitational
potentials.\par\medskip 
This classical approach based on deformation theory, differs
from the classical gauge one in general relativity
\cite{ivan}. Indeed  the latter are developed out of a given
gauge Lie group. But at first, they are not Lie pseudogroups,
and in a second place,  they are assossiated to Lie groups 
invariance of the tangent spaces (not the tangent fiber
bundle but the fibers) at any fixed base point $p_0$. They
can accordingly be regarded as isotropy Lie
subgroups of the corresponding pseudogroups. For instance, in
fixing the function
$\alpha$ to a constant, the set of
applications $\hf$ becomes a Lie group and not a Lie pseudogroup. In
that case the applications $\hf$ would depend on 15 real
variables at $n=4$, and no longer on a  set of arbitrary
functions, as we will be seen in the conformal
pseudogroup case.\par\medskip  Also it is neither a
Kaluza-Klein type theory nor is it based on a Riemann-Cartan
geometry. In the present model, there is no torsion.  It also
completly differs from the H. Weyl unifications  and the
J.-M. Souriau approach
\cite{souriau,weylbook,weylOfar}. 
Close approaches to ours, are
developed on the one hand by M. O. Katanaev \& I. V. Volovich
\cite{katavol92}, and on the other hand by H. Kleinert \cite{kleinbook2}
and J.-F. Pommaret \cite{pommbook94}. At lower
dimensions, other general relativity models
are investigated within similar approaches such as models of
gravity   in \cite{deserjackiwhooft,giddingsabbottkuchar,gottalplert}
for instance. In fact, our present work can somehow be
viewed as an extension of the T. Fulton
\textit{et al.} approach and model \cite{fulton62}, or, as
a continuation of the original works of J.
Haantjes \cite{haantjes41}.\par\smallskip {\em Here below, we
summarize the mathematical procedure and assumptions  presented in
the sequel and based on the previous discussion (we refer to
definitions of  involution i.e. integrability, symbols of
differential equations, acyclicity and formal integrability such as
those given in 
\cite{bryantal,gasquigold,guillstenb66,pommbook94,spencerlin69} for
instance):
\begin{enumerate}
\item[$\bullet$] The metric field $\omega$ is conformally equivariant.
\item[$\bullet$] The Riemann scalar curvature $\rho_s$ associated
to the metric field $\omega$, is assumed to be a constant,
$n(n-1)k_0$, as a
consequence of the constant Riemann curvature tensor
assumption. And then, the Weyl tensor associated to
$\omega$ is vanishing,
i.e., we have a conformally flat structure.
\item[$\bullet$] The  system (\ref{first}) of differential equations
in $\hf$, being  non-integrable, will be
supplied with an other  system of equations, obtained from a
prolongation procedure which will be stopped as the integrability
conditions of the  resulting complete system of partial differential
equations is met.
\item[$\bullet$] The covariant derivatives involved in the
prolongation procedure will be assumed to be torsion free, i.e. we
will make use of the Levi-Civita covariant derivatives.
\item[$\bullet$] We will extract from the latter system of PDE,
a subsystem, which will be called the ``Generalized
Haantjes-Schouten-Struik (GHSS) system" (see system
(\ref{third}) \cite[with $k_0=0$]{haantjes41,schoutstruik38}), defining completely the
sub-pseudogroup
$\SGammaH$ of those applications
$\hf$ which are strictly smooth deformations in $\GammaG$ of 
applications
$f\in\GammaP$. This PDE subsystem   will be satisfied by the
functions
$\alpha$. {\em This is the core system of our model and to our
knowledge it has never been really studied, or, at least,
related to any unification  model. The S-admissibility
property appears at this step as a system of PDE satisfied by
the metric field
$\omega$; A system which strongly resembles an
Einstein equation.\/}
\item[$\bullet$] By considering Taylor series  solutions to
the ``GHSS system", we will show how  general solutions
depend on a particular finite set of parametrizing functions.
\item[$\bullet$] We show that this set of parametrizing functions is
associated to a Spencer differential sequence \cite{spencerlin69},
and that they can be identified with  both electromagnetic and
gravitational gauge potentials.
\item[$\bullet$] We deduce the  metric field
$\nu\equiv\omega+\delta\omega$ of the infinitesimal smooth
deformations, depending on the electromagnetic and gravitational
gauge potentials, from which covariant derivatives would be deduced.
The Newtonian limit will also be indicated as well as the meaning of
the so-called \emph{``meshing assumption"\/} \cite{ghins} in the
present context.
\end{enumerate}}
To finish, we indicate that the mathematical tools
used for this unification  finds its roots, first in  the
conformal Lie structure that has  been extensively
studied by H. Weyl \cite{weylkonf}, K. Yano \cite{yanobook70},
J.~Gasqui \cite{gasqui}, J. Gasqui
\& H. Goldschmidt \cite{gasquigold} and J.-F.~Pommaret
\cite{pommbook94}. Meanwhile, we only partially refer to some of
these aspects since it mainly has to do with the general theory of
Lie equations, and not exactly with the set of PDE we are concerned
with. We essentially indicate,
succintly, the cornerstones which are absolutely necessary for our
explanations and descriptions of the present framework.

%%%%%%%%%%%%%%%%%%%%%%%%%%%%%%%%%%%%%%%%%%%%%%%%%%%%%%
%%%%%%%%%%%%%%%%%%%%%%%%%%%%%%%%%%%%%%%%%%%%%%%%%%%%%%
%%%%%%%%%%%%%%%%%%%%%%%%%%%%%%%%%%%%%%%%%%%%%%%%%%%%%%

\section{The conformal finite Lie equations of the substratum spacetime}
First of all, and from the previous sections, we assume that
the pseudogroup of relativity is no longer  Poincar\'{e}  but
the conformal Lie pseudogroup. In particular, this means that no
physical law changes occur, shifting from a given frame
embedded in a gravitational field,  to a uniformly accelerated
relative isolated one \cite{pageI,pageII}.\par The conformal finite
Lie equations are deduced from the conformal action on a {\it
local\/} me\-tric field
$\omega$ defining a  pseudo-Riemannian structure on
$\Rset^n\simeq\Sub$. We insist on the fact we do local studies, 
meaning that we consider {\it local\/}
charts from open subsets of the latter manifolds into a
common open subset of $\Rset^n$. Hence by geometric objects
or computations on $\Rset^n$, we mean {\it local\/} geometric
objects or computations  on the
manifolds $\m$, $T\m$ and/or $\Sub$. Also, it is well-known that the
mathematical results displayed below are independent of the dimension
when the latter is greater or equal to 4 \cite{gasquigold}.\par Let
us consider
${\hf}\in\diff{\infty}(\Rset^n)$, the set of local
diffeomorphisms of $\Rset^n$ of class $C^\infty$, and any
function $\alpha\in{C^\infty}(\Rset^n,{\Rset})$. Then if ${\hf}\in{\GammaG}$
($\GammaG$ being the pseudogroup of local conformal bidifferential
maps on $\Rset^n$), ${\hf}$ is a solution  of the PDE system
(\ref{first}). In fact other PDE must be satisfied to completely
define $\GammaG$ as will be seen in the sequel. Also, only
the $+\e^{2\alpha}$ positive conformal factors are  retained so as to
preserve  one orientation only on $\Rset^n$, and we will accordingly restrict ourselves
to those $\hf$ which preserve that
orientation (we  recall that $\alpha$ {\em is a varying
function depending on each $\hf$ and consequently not
fixed\/}). We denote
$\tilde{\omega}_f$ the metric on $\Rset^n$ such that by
definition: $\hf^{\ast}(\omega)\equiv\tilde{\omega}_f$, and we
agree on putting a tilde on each tensor or geometrical
``object" relative to, or deduced from the  metric
$\tilde{\omega}_f$. 
\par Now, performing a first prolongation of the system
(\ref{first}), we deduce another second order system of PDE's
connecting the Levi-Civita covariant derivatives  $\nabla$ and
$\widetilde{\nabla}$, respectively associated to $\omega$
and $\tilde{\omega}_f$. These new  differential equations are 
(see for instance \cite{gasqui}) $\forall\, X,\,Y\in
T\Rset^n$:
%%%%%%%%%%%%%%%%%%%%%%%%%%%%%%%%%%%%%%%%%%%%%% 
\begin{equation}
{{\widetilde{\nabla}}_X}Y={\nabla_X}Y+d\alpha(X)Y+d\alpha(Y)X-
\omega(X,Y)\,{{}_\ast}d\alpha\,, 
\label{3} 
\end{equation}
%%%%%%%%%%%%%%%%%%%%%%%%%%%%%%%%%%%%%%%%%%%%%%%%%%%%%%%%%%%%
where $d$ is the exterior differential and ${{}_\ast}d\alpha$
is the ``$\omega$-dual" vector field of the 1-form $d\alpha$
with respect to the metric $\omega$, i.e., such that $\forall\,
X\in T\;\Rset^n$:
%%%%%%%%%%%%%%%%%%%%%%%%%%%%%%%%%%%%%%%%%%%%%% 
\begin{equation}
\omega({{}_\ast}d\alpha,X)=d\alpha(X)=<d\alpha|X>\,.
\label{4} \end{equation}
%%%%%%%%%%%%%%%%%%%%%%%%%%%%%%%%%%%%%%%%%%%%%%%%%%%%%%%%%%%
Since the Weyl tensor $\tau$ associated to $\omega$ is
vanishing, the Riemann tensor $\rho$ can be
rewritten
$\forall\,X,Y,Z,U\in{C^\infty}(T\Rset^n)$ as:
%%%%%%%%%%%%%%%%%%%%% Equation7 %%%%%%%%%%%%%%%%%%%%%%%%%%
\begin{eqnarray} 
\omega(U,\rho(X,Y)Z)&=&{\frac{1}{(n-2)}}
\left\{\omega(X,U)\sigma(Y,Z)-
\omega(Y,U)\sigma(X,Z)\right.\nonumber\\
%\hspace{4cm}
&&\left.+\,\omega(Y,Z)\sigma(X,U)-
\omega(X,Z)\sigma(Y,U)\right\}, 
\label{6} 
\end{eqnarray}
%%%%%%%%%%%%%%%%%%%%%%%%%%%%%%%%%%%%%%%%%%%%%%%%%%%%%%%%%%% 
where $\sigma$ is  defined  by (see the tensor ``\,L\," in \cite{yanobook70} up
to a constant depending on $n$)
%%%%%%%%%%%%%%%%%%% Definition de sigma %%%%%%%%%%%%%%%%%%%
\begin{equation}
\sigma(X,Y)={\rho_{\mathrm{ic}}}(X,Y)-
{\frac{\rho_{\mathrm{s}}}{2(n-1)}}\,\omega(X,Y),
\label{7} 
\end{equation}
%%%%%%%%%%%%%%%%%%%%%%%%%%%%%%%%%%%%%%%%%%%%%%%%%%%%%%%%%%% 
where $\rho_{\mathrm{ic}}$ is the Ricci tensor and $\rho_{\mathrm{s}}$
is the Riemann scalar curvature. Consequently, the  first
order system of PDE in $\hat{f}$ ``connecting" $\tilde{\rho}$
and $\rho$, can be rewritten as a first order system of PDE
concerning $\tilde{\sigma}$ and
$\sigma$. Using the torsion
free property of the Levi-Civita covariant derivatives and applying
again the covariant derivative $\widetilde{\nabla}$ on the relation
(\ref{3}), one obtains the following
\textit{third\/} order system of PDE (since $\alpha$ is depending on the first
order derivatives of $\hf$):
%%%%%%%%%%%%%%%%%%%%%%%%%%%%%%%%%%%%%%%%%%%%%%%%%%%%% 
\begin{eqnarray} 
\hf^\ast(\sigma)(X,Y)\equiv\tilde{\sigma}(X,Y)&=&
\sigma(X,Y)+(n-2)\Big(d\alpha(X)
d\alpha(Y)\nonumber\\
%\hspace{5cm}
&&-{\frac{1}{2}}\omega(X,Y)
d\alpha({{}_\ast}
d\alpha)
-\mu(X,Y)\Big),
\label{10} 
\end{eqnarray}
in which we have defined the symmetric
tensor $\mu\in{C^\infty}({S^2}\Rset^n)$ by:
%%%%%%%%%%%%%%%%%%%%%%%%%%%%%%%%%%%%%%%%%%%%%%%%%%%%%%%%%%%%%%%%%
\begin{equation}
\mu(X,Y)=\frac{1}{2}\left\{<\nabla_X(d\alpha)|Y>+
<\nabla_Y(d\alpha)|X>\right\}\,.
\label{9}
\end{equation}
%%%%%%%%%%%%%%%%%%%%%%%%%%%%%%%%%%%%%%%%%%%%%%%%%%%%%%%%%%%%%%%%%
To go further, it is important  again to notice that the relation
(\ref{10}) is directly related to a third order system of
PDE, we will denote by (T), since it is deduced from a
supplementary prolongation procedure applied to the second
order system  (\ref{3}). Then it follows, from  the
well-known theorem of H. Weyl on the equivalence of conformal
structures
\cite{gasquigold,weylkonf,yanobook70}, and because of the
Weyl tensor vanishing, that the systems of differential
equations (\ref{first}) and (\ref{3})  when completed
with the latter third order system (T), becomes an
involutive system of order three. Let us stress again
that $\alpha$ is  merely  defined by $\hf$ and its first order
derivatives, according to the relation (\ref{first}).\par Looking
only at those applications $\hf$  which are smooth deployments
of applications $f$,  this third order system of PDE must
reduce to  a particular  system of PDE, associated
to a conformal Lie sub-pseudogroup we denote by $\SGammaH$.
Indeed, if
$\alpha$ tends towards the zero function (with respect to the
$C^2$-topology) then, the previous set of smoothly deformed applications
$\hf$ must tend towards the Poincar{\'e} Lie pseudogroup. But this
condition is not satisfied by all of the applications $\hf$
in the conformal Lie pseudogroup $\GammaG$, since in full
generality, the non-trivial third order system of PDE (T)
would be kept at the zero
$\alpha$ function limit: the sub-pseudogroup
$\SGammaH$ would have to be defined by an involutive second
order system of PDE which would tend towards the involutive
system defining the Poincar{\'e} Lie pseudogroup.\par The
systems of differential equations (\ref{first}) and (\ref{3})
would be well suited to define partially this pseudogroup
$\SGammaH$. The
$n$-acyclicity property of $\SGammaH$ would be restored and borrowed,
at the order two, from the  Poincar{\'e} one, provided however that  an,
a priori, arbitrary  \emph{input\/} perturbative function $\alpha$ is
given before, instead of being defined from an application $\hf$
in accordance with the relation (\ref{first}). But the formal integrability
is obtained  only if  the tensor $\sigma$  satisfies one of the
following  equivalent relations (see fromula (16.3) with definition
(3.12) in 
\cite{gasquigold}):
%%%%%%%%%%%%%%%%%%%%%%%%%%%%%%%%%%%%%%%%%%%%%%%%%%%%%%%%%%%%%%%%%
\begin{equation} 
\sigma={k_0}{\frac{(n-2)}{2}}\,\omega\quad
\Longleftrightarrow\quad\rho_{\mathrm{ic}}=(n-1)\,k_0\,\omega\,,
\label{11}
\end{equation}
%%%%%%%%%%%%%%%%%%%%%%%%%%%%%%%%%%%%%%%%%%%%%%%%%%%%%%%%%%%%%%%%%
deduced from relation (\ref{7}), in order to avoid adding up
the supplementary  first order system of PDE (\ref{10}) to
(\ref{first}), $\alpha$ being considered an input
function. In fact, as it is well-known, the
relations above are a consequences of the constant Riemann
curvature tensor assumption, but they appear in a different
way. Then, considering the system (\ref{first}) and
(\ref{11}), the system (\ref{10}) reduces to  a second order
system of PDE  concerning the input function
$\alpha$ only, which is thus constrained, contrarily to what
might have been expected, and such that:
%%%%%%%%%%%%%%%%%%%%%%%%%%%%%%%%%%%%%%%%%%%%%%%%%%%%% 
\begin{equation}
\mu(X,Y)=
{\frac{1}{2}}\left\{
\left[{k_0}\left(1-{\e^{2\alpha}}\right)
-d\alpha({{}_\ast}d\alpha)
\right]\omega(X,Y)\right\}+d\alpha(X)d\alpha(Y)\,. 
\label{12} 
\end{equation}
%%%%%%%%%%%%%%%%%%%%%%%%%%%%%%%%%%%%%%%%%%%%%%%%%%%%%%%%%%%%%%%%%
Obviously, as can be easily verified, this is an involutive system of
PDE, since it  is a formally integrable system with an elliptic
 symbol (i.e. a vanishing  symbol) of order two.\par
Thus, we have  series of PDE deduced from (\ref{first})
defining all the
\emph{smooth\/} deformations of the applications $f$ contained in the
conformal Lie pseudogroup $\GammaG$.\par In addition, the metric field
$\omega$ satisfies the relation
(\ref{11}) (which is an Einstein equation when the
stress-energy tensor is proportional to the metric, or when
the latter is vanishing but with a non-zero  cosmological
term). In that case, such a metric field
$\omega$, or the substratum spacetime $\Sub$ will be called
\textit{S-admissible} (with S as ``Substratum"),
and this S-admissibility is assumed in the
sequel.\par   Setting
\(
\omega=\sum_{i,j=1}^{n}\omega_{ij}(x)\,dx^i\otimes\,dx^j\,,
\)
in an orthonormal
system of coordinates, then the PDE
(\ref{first}), (\ref{3}) and (\ref{12}) defining
$\SGammaH\subset\GammaG$ can be written, with
$\det(J({\hf}))\not= 0$ and $i,j,k=1,\cdots,n$ as:
\begin{subequations}
\begin{eqnarray}
&&\sum_{r,s=1}^n\omega_{rs}(\hf)\,\,\hf^{r}_{i}\hf^{s}_{j}=
e^{2\alpha}\omega_{ij}\,,
\label{f1}\\ 
&&\hf^k_{ij}+
\sum_{r,s=1}^n{}\gamma^{k}_{rs}(
\hf)\,\,\hf^{r}_{i}\hf^{s}_{j}=
\sum_{q=1}^n{}\hf_q^k\left(
\gamma^{q}_{ij}+\alpha_i\delta^q_j+
\alpha_j\delta^q_i-\omega_{ij}\alpha^q\right)\,,
\label{f2}\\
&&\mu_{ij}=\alpha_{ij}-
\sum_{k=1}^n\alpha_k\gamma^k_{ij}=
\frac{1}{2}
\Big\{k_0(1-\mbox{e}^{2\alpha})-
\sum_{k=1}^n\alpha^k\alpha_k
\,\Big\}\omega_{ij}+
\alpha_i\alpha_j\,,\label{third}
\end{eqnarray}
\label{confeq}
\end{subequations}\par\noindent
where $\delta^i_j$ is the Kronecker tensor, and where one
denotes as usual
 $\hf^i_j\equiv\partial \hf^i/\partial
x^j\equiv\partial_j \hf^i$, etc~\dots, $T_k=\sum_{h=1}^n
T^h\,\omega_{hk}$ and $T^k=\sum_{h=1}^n
T_h\,\omega^{hk}$ for any tensor $T$, where $\omega^{ij}$
is the inverse metric tensor, and
$\gamma$ is the  Riemann-Christoffel form associated to the
\emph{S-admissible\/} metric $\omega$. This is the set of our
starting equations. It matters to notice that the (T) system is not
included in the above set of PDE. Indeed, the latter being
already involutive from order 2, this involves, by
definition of involution, that the (T) system is redundant
since all the applications
$\hf\in\SGammaH$, solutions  of (\ref{confeq}), will also
be solutions of all the  systems of PDE obtained by prolongation.
\par It is pertinent to notice that
$\mu$ or, equivalently, the tensor
$\tilde{\alpha}_2\equiv\{\alpha_{ij},i,j=1,\cdots,n\}$ might be
considered as an Abraham-E\"{o}tv\"os type tensor
\cite{abraham,misner} encountered in the  E\"otv\"os-Dicke
experiments for the measurement of the stress-energy tensor of the
gravitational potential.\par At this point, it may also be
worth  making contact with some previous set 
of  physical interpretations
\cite{haantjes40,haantjes41,pageI,pageII,schoutstruik38} (up to a constant
for units and with
$n=4$) according to which  the tensor
$\mu$ is ascribed to the stress-energy tensor,
$\tilde{\alpha}_1$ to the gravity acceleration 4-vector,  and
$\alpha$ to the Newtonian potential of gravitation.
%%%%%%%%%%%%%%%%%%%%%%%%%%%%%%%%%%%%%%%%%%%%%%%%%%%%%%
%%%%%%%%%%%%%%%%%%%%%%%%%%%%%%%%%%%%%%%%%%%%%%%%%%%%%%
%%%%%%%%%%%%%%%%%%%%%%%%%%%%%%%%%%%%%%%%%%%%%%%%%%%%%%

\section{Functional dependence of the spacetime deployment}
Let us now look  for the formal series, solutions of the PDE
system (\ref{confeq}), assuming from now on, that the
metric $\omega$ is  analytic. We know these series will be
convergent in a suitable open subset and thus will provide
analytic solutions,  since the analytic system is involutive
and in particular elliptic, because of a vanishing symbol
(see, in appendix 4 of
\cite{malgrangeII}, the Malgrange theorem for elliptic
systems). Nevertheless, we need of course to know the Taylor
coefficients. For instance, we can choose  for the
applications
$\hat{f}$ and the functions
$\alpha$  the following series at a point
$x_0\in\Rset^n$:
\begin{eqnarray}
\hf^i(x):\,
S^i(x,x_0,\{\hat a\})=\sum_{|J|\geq0}^{+\infty}
\hat{a}^i_J(x-x_0)^J/|J|!\,,
\label{serief}\\
\alpha(x):\,
s(x,x_0,\{c\})=\sum_{|K|\geq0}^{+\infty} 
c_K(x-x_0)^K/|K|!\,,\nonumber
\end{eqnarray}
with $x\in U({x_0})\subset\Rset^n$ being a suitable open 
neighborhood of $x_0$ to insure the convergence of
the series,
$i=1,\cdots,n$, $J$ and
$K$ are multiple index notations such as $J=(j_1,\cdots,j_n)$, 
$K=(k_1,\cdots,k_n)$ with $|J|=\sum_{i=1}^n j_i$ and
similar expressions for $|K|$. Likewise, $\{\hat{a}\}$ and
$\{c\}$ are  sets of Taylor coefficients, whereas the
$\hat{a}^i_J$ and $c_K$ are real values and not functions
of $x_0$, though of course, they can also be values of functions at
$x_0$ (let us remark that we could  consider instead the
vector: $\xi=x-x_0\in\tom$, which strengthens the equivalence
principle we are using with $\tom$ as substratum spacetime
$\Sub$).
\par Also we must add that usual partial derivatives will
be used in these Taylor series determinations instead of
covariant derivatives. The use of either of the two derivatives is
thoroughly as discussed in the appendix, and shown to be basically
equivalent.
%%%%%%%%%%%%%%%%%%%%%%%%%%%%%%%%%%%%%%%%%%%%%%%%%%%%%%
%%%%%%%%%%%%%%%%%%%%%%%%%%%%%%%%%%%%%%%%%%%%%%%%%%%%%%

\subsection{The ``Generalized Haantjes-Schouten-Struik system"}
We call the ``Generalized Haantjes-Schouten-Struik system" (GHSS
system), the system of PDE (\ref{third}) (see
\cite{fulton62,haantjes41,schoutstruik38} for this system,  but at
$k_0=0$). It is from this set of PDE that  gauge potentials and fields of
interactions could occur. From the series $s$, at zero-th order one
obtains the\textit{ algebraic equations\/} ($i,j=1,\cdots,n$;
$\bfc_1=\{c_1,\cdots,c_n\}$):
\begin{eqnarray}
c_{ij}&=&\frac{1}{2}\Big\{k_0(1-\e^{2c_0})-
\sum_{k,h=1}^n\omega^{kh}(x_0)c_h c_k
\,\Big\}\omega_{ij}(x_0)+c_ic_j+\sum_{k=1}^n\,c_k\gamma^k_{ij}\nonumber\\
&\equiv&
F_{ij}(x_0,c_0,\bfc_1)\,,
\label{c2eq}
\end{eqnarray}
and it follows that the $c_K$'s such that $|K|\geq2$, will
depend recursively only on $x_0$, $c_0$ and $\bfc_1$.
It is none but the least the meaning of formal integrability of
so-called involutive systems. Hence the
series for $\alpha$ can be written as a convergent series,
$s(x,x_0,c_0,\bfc_1)$, developed with respect to powers of
$(x-x_0)$, $c_0$ and $\bfc_1$. Let us notice that we can change or not the
values at $x$ of the series $s$, by varying $x_0$, $c_0$ or
$\bfc_1$.\par
Let $J_1$ be the 1-jets affine bundle of the
$C^\infty$ real valued functions on $\Rset^n$. Then, from the
latter remark, it exists a subset associated to
$\bfc_0^1\equiv(x_0,c_0,\bfc_1)\in J_1$, we denote by
$\Su(\bfc_0^1)\subset J_1$, the set of elements
$({x}_0',{c}_0',\bfc_1')\in J_1$, such that there
is an open neighborhood
$U(\bfc_0^1)\subset \Su(\bfc_0^1)$, being projected on $\Rset^n$ in an open
neighborhood of a given
$x\in U(x_0)$, for which, for all $({x}_0',{c}_0',\bfc_1')\in
U(\bfc_0^1)$, then,
$s(x,x_0,c_0,\bfc_1)=s(x,{x}_0',{c}_0',\bfc_1')$.
Assuming that the variation $ds$ with respect
to $x_0$, $c_0$ and $\bfc_1$ is vanishing, at a given fixed $x$, is the
subset $\Su(\bfc_0^1)$ a submanifold of $J_1$ ? From $ds\equiv0$ it
follows that ($k=1,\cdots,n$):
\begin{eqnarray*}
&&\sg_0\equiv dc_0-\sum_{i=1}^nc_i\,dx_0^i=0\,,
\\
&&\sg_k\equiv dc_k-
\sum_{j=1}^n F_{kj}(x_0,c_0,\bfc_1)\,dx_0^j=0\,.
\end{eqnarray*}
We recognize a regular analytic Pfaff system,  we denote by
$P_c$, generated by the 1-forms $\sg_0$ and $\sg_k$, and
the meaning of their vanishing is that the solutions
$s$ for $\alpha$ do not change for such variations of $c_0$,
$\bfc_1$ and $x_0$. Also, as can be easily
verified, the Pfaff system $P_c$ is integrable since the
Fr\"obenius conditions of involution are satisfied, and all of the
prolongated 1-forms $\sg_K$ with $|K|\geq2$, are linear
combinations of these $n+1$ generating forms, thanks to the
recursion property of formal integrability. Then the subset
$\Su(\bfc_0^1)$ of dimension $n$ containing a particular
element $\bfc_0^1\equiv(x_0,c_0,\bfc_1)$, is a submanifold of
$J_1$. It is a particular leaf of, at least, a local foliation $\mathcal{F}_1$ on $J_1$
of codimension $n+1$.\par\medskip
Then, since the system of
PDE defined by the involutive Pfaff system
$P_c$, namely the GHSS system, is elliptic (i.e., vanishing 
symbol in the present case)
and formally integrable, one deduces that it exists on
$J_1$, local analytic systems of coordinates
$(\tau_0,\tau_1,\cdots,\tau_n)$ of the transverse
submanifold of the foliation, such that each leaf
$\Su(\bfc_0^1)$ is an analytic submanifold
\cite{malgrangeII} for which $\tau_0=cst$ and $\tau_i=cst$ $(i=1,\cdots
,n)$. In other words, all the  series
$s(x,{x}_0',{c}_0',\bfc_1')$ with
$({x}_0',{c}_0',\bfc_1')\in\Su(\bfc_0^1)$ are convergent
and correspond to one single analytic solution 
$u(x,\tau_0,\boldsymbol{\tau}_1)$ 
($\boldsymbol{\tau}_1\equiv\{\tau_1,\cdots,\tau_n\}$),
analytic  with respect to $x$  as well as  with respect to the
$\tau$'s. This results from  the $s$ continuous series
convergent character, whatever the fixed set of given values
$x$, $x_0$, $c_0$  and $\bfc_1$. Thus, in full generality,
considering the difference
$s(x,x_0,c_0,\bfc_1)-s(x,{x}_0',{c}_0',\bfc_1')$
we have the relation:
\begin{equation}
s(x,x_0,c_0,\bfc_1)-s(x,{x}_0',{c}_0',\bfc_1')=
u(x,\tau_0,\boldsymbol{\tau}_1)-
u(x,{\tau}_0',\boldsymbol{\tau}_1')\,,
\label{diffs}
\end{equation}
with the $\tau$ parameters related by ($i=1,\cdots,n$)
\begin{equation}
{\tau}_0'-\tau_0=
\int_{\bfc^1_0}^{{\bfc'}^1_0}\!
\sg_0\,,\qquad{\tau}_i'-\tau_i=
\int_{\bfc^1_0}^{{\bfc'}^1_0}\!
\sg_i\,.
\label{integsig}
\end{equation}
Now, we consider the
$c$'s as values of differential (i.e. $C^\infty$) functions
$\rho\,$: $c_K=\rho_K(x_0)$, as expected for usual Taylor series
coefficients, and defined on a starlike open neighborhood of 
$x_0$ (the integrals above define none but the least than  a
homotopy operator, and that ``starlike" open subsets obviously mean 
simply connected open subsets 
\cite{buttin67,spencerlin69}).
Roughly speaking, we make a pull-back on
$\Rset^n$ by differentiable sections $\rho$, inducing a projection
from the subbundle of projectable elements in
$T^\ast\!J_1$ to
$T^\ast\Rset^n\otimes_\Rset\!J_1$. Then, we set 
(with $\boldsymbol{\rho}_1\equiv\{\rho_1,\cdots,\rho_n\}$,
$\partial^0_i\equiv\partial\,/\partial x_0^i$
and no changes of notations for the pull-backs):
\begin{subequations}
\begin{eqnarray}
&&\sg_0\equiv\sum_{i=1}^n(\partial^0_i\rho_0-\rho_i)\,dx_0^i\equiv
\sum_{i=1}^n\A_i\,dx_0^i\,,
\label{rhoA}\\
&&\sg_i\equiv\sum_{j=1}^n(\partial^0_j\rho_i-
F_{ij}(x_0,\rho_0,\boldsymbol{\rho}_1))\,dx_0^j
\equiv\sum_{j=1}^n\B_{j,i}\,dx_0^j\,,
\end{eqnarray}
\label{rhoAB}
\end{subequations}\par\noindent
and it follows that the integrals (\ref{integsig}) must be
performed from $x_0$ to ${x}_0'$ in a starlike open neighborhood of
$x_0$. In particular, if
$\bfc_0^1$ is an element of  the ``null" submanifold, or
stratum, corresponding to the vanishing solution of the ``GHSS
system", then  the difference (\ref{diffs}) involves that
\[
\alpha(x)\equiv
s(x,{x}_0',{c}_0',\bfc_1')=
u(x,{\tau}_0',\boldsymbol{\tau}_1')\,,
\]
with
\[
{\tau}_0'=
\int_{x_0}^{{x}_0'}
\sum_{i=1}^n\A_i\,dx^i+\tau_0\,,\qquad{\tau}_i'=
\int_{x_0}^{{x}_0'}
\sum_{j=1}^n\B_{j,i}\,dx^j+\tau_i\,.
\]
This result displays the functional dependencies of the $\tau$ deformation parameters 
of the solutions of the ``GHSS system", with respect to the functions $\rho_0$
and $\boldsymbol{\rho}_1$.
These  smooth infinitesimal deformations define  the fields $\A$
and $\B$, i.e., $n(n+1)$ potential functions ($\bf 20$
functions if $n=4$), which can also be considered as infinitesimal smooth
deployments from ``Poincar{\'e} solutions" of the system
(\ref{confeq}) at $\alpha\equiv0$, to some ``conformal
solutions" whatever is
$\alpha$ satisfying the GHSS system.\par Moreover the
functions
$\rho$, and consequently the fields $\A$ and
$\B$, must satisfy additional differential equations
coming from  the Fr\"obenius conditions of involution for the
Pfaff system $P_c$. More precisely, from the relations
$d\sg_0=\sum_{i=1}^n\,dx_0^i\wedge\sg_i$,
$d\sg_i=\sum_{j=1}^n\,dx_0^j\wedge\sg_{ij}$ and 
\begin{eqnarray}
\sg_{ij}&=&c_i\sg_j+c_j\sg_i-\omega_{ij}\Big\{k_0\e^{2c_0}\sg_0+
\sum_{k,h=1}^n\omega^{kh}c_h\sg_k\Big\}
+\sum_{k=1}^{n}\gamma^k_{ij}\sigma_k\nonumber\\
&\equiv&\vartheta_{ij}(\bfc_0^1,\sg_J;|J|\leq1)\,,
\label{sg2}
\end{eqnarray}
one deduces
a set of algebraic relations to be satisfied at $x_0$:
\begin{subequations}
\begin{eqnarray}
\J_{k,j,i}&\equiv&
\omega_{ij}\left\{k_0e^{2\rho_0}\A_k+
\sum_{r,s=1}^n
\omega^{rs}\,\rho_r\,\B_{k,s}\right\}
\nonumber\\
&&\hspace{1cm}+\partial^0_j\B_{k,i}
-\rho_i\,\B_{k,j}-
\rho_j\,\B_{k,i}
-\sum_{s=1}^{n}\gamma^s_{ij}\B_{k,s}\,,
\label{JAB}
\end{eqnarray}
\begin{equation}
%\fl
\I_{i,k}=\I_{k,i}\equiv\partial^0_k\A_i-\B_{i,k}\,,
\qquad \J_{k,j,i}=\J_{j,k,i}\,.\label{IAB}
\end{equation}
\label{ABIJeq}
\end{subequations}  
In these relations, the set of functions
$(\rho_0,\boldsymbol{\rho}_1)$ appears to be, \textit{a priori\/} only, a set of
arbitrary differential functions. Finally, we deduce:
\begin{theorem}
All the analytic solutions of the involutive system of PDE\,
(\ref{third}) can be written in a suitable starlike open
neighborhood $U(x_0)$ of $x_0$ as
\begin{eqnarray}
&\alpha(x)\equiv u\Big(x,\int_{x_0}^{{x}_0'}
\sum_{i=1}^n\A_i\,{dx'}^i+\tau_0\,,\int_{x_0}^{{x}_0'}
\sum_{j=1}^n\B_{j,1}\,{dx'}^j+\tau_1
,\cdots\nonumber\\
&\hspace{7cm}
,\int_{x_0}^{{x}_0'}
\sum_{j=1}^n\B_{j,n}\,{dx'}^j+\tau_n\,\Big)\,,
\label{deformpar}
\end{eqnarray}
with
\(u(x_0,\tau_0,\tau_1,\cdots,\tau_n)=0\), $x'_0\in U(x_0)$, and
where $u$ is a unique fixed analytic function depending on
the $\mbox{n(n+1)}$ $C^\infty$
integrable functions
$\A_i$ and $\B_{j,k}$ defined by the relations (\ref{rhoAB}).
The integrals in $u$ are called the ``potential of
interactions". Let us remark that we can set $x'_0\equiv
v(x)$ if the gradient of $v$, i.e. $\nabla v$, is
in the annihilator of the Pfaff system $P_c$ of
1-forms $\sigma$.
\end{theorem}
This Theorem making explicit the dependence of the GHSS system solutions on the $\A$
and $\B$ ``gauge fields", can be viewed as an illustration of the well-known
Cartan-K\"ahler Theorem, and indicates also  how the fields $\A$ and $\B$ ``gauge"
the geometrical deformations of the (isometries) Poincar{\'e} Lie pseudogroup.\par
Also,  considering physical aspects, and defining 
$\F$ and $\G$ as being the respectively  skew-symmetric and 
symmetric parts of the tensor of components
$\partial_i\rho_j$, one deduces, from the symmetry
properties of the  relations (\ref{ABIJeq}), what we call {\it the
first set of differential equations associated to $\Sub$} at $x_0$:
\begin{subequations}
\begin{eqnarray}
&&\partial^0_i\F_{jk}+\partial^0_j\F_{ki}+\partial^0_k\F_{ij}=0\,,
\label{max1}\\
&&2\,\partial^0_j\G_{ki}-\partial^0_i\G_{kj}-\partial^0_k\G_{ij}=
\partial^0_i\F_{jk}-\partial^0_k\F_{ij}\,,
\end{eqnarray}
\label{firstset}
\end{subequations}
with  
\begin{subequations}
\begin{eqnarray}
&&\F_{ij}=\partial^0_j\rho_i-\partial^0_i\rho_j
=\partial^0_i\A_j-\partial^0_j\A_i\,,
\label{set1}\\
&&\G_{ij}=-(\partial^0_i\rho_j+\partial^0_j\rho_i)\equiv
\partial^0_i\A_j+\partial^0_j\A_i
\mod(\rho_0,\partial_i{\rho}_0)\,.
\end{eqnarray}
\label{FG}
\end{subequations}\par\noindent
The PDE (\ref{max1}) with (\ref{set1})  can be interpreted as the
first set of {\it Maxwell equations}. In view of physical
interpretations, we can easily  compute the Euler-Lagrange equations
of a conformally equivariant Lagrangian density
\begin{subequations}
\begin{equation}
\Lag(x_0,\rho_0,
\boldsymbol{\rho}_1,\A,\F,\G)\,d^n\!x_0\,,
\label{LAB}
\end{equation}
or more generally
\begin{equation}
\Lag(x_0,\rho_0,\boldsymbol{\rho}_1,\A,\B,\I,\J)\,d^n\!x_0\,,
\label{LABIJ}
\end{equation}
\label{LAGDENS}
\end{subequations}
with $\A$, $\B$, $\F$, $\G$, $\I$ and $\J$ satisfying the relations
(\ref{rhoAB}), (\ref{FG}) and (\ref{ABIJeq}). Doing so, we would obtain easily
what could be dubbed {\it the second set of differential equations
associated to $\Sub$} at $x_0$, some aspects of which will be discussed in the last
section.
\par\smallskip Then to proceed further, a few well-known definitions are in
order \cite{atimacdo69}. We call
\textit{germ} at $x_0$ of an application $f$, the class of  
$C^{\infty}$ applications $\tilde f$,  for which it exist an 
open subset $U$ of $x_0$ such that $f/_U=\tilde f/_U$. We call  \textit{ring\/} on $U$
or \textit{local ring\/} (if it contains a unique maximal proper ideal) on $U$ a set of germs of
applications defined on
$U$, all satisfying possibly the same given ``\textit{formulae\/}" on a given subset of
points in $U$ (i.e., on a
\textit{ponctuated} open subset $U$). The set of rings can be
endowed with  a so-called 
\textit{presheaf} structure. In the sequel, we will not consider
sheafs, since the compatibility condition, defining sheafs
from presheafs, will not be used here because only local (on
one given open subset) topological considerations will be
relevant to our concern.
\par
\begin{definition} We denote:\par
\begin{enumerate}
\item $\theta_\Rset$, the presheaf of rings of germs of
the differential (i.e.
$C^\infty$) functions defined on $\Rset^n$, 
\item $\underline{J_1}$, the presheaf  of
$\theta_\Rset$-modules of germs of differential sections of
$J_1$,
\item $\Sz\subset \theta_\Rset$, the presheaf  of rings
of germs of functions which are solutions with their first derivatives, of the 
``algebraic equations" GHSS (\ref{third}) taken at any given points
$x_0$ in $\Rset^n$, \underline{not simultaneously} at each point in $\Rset^n$ (see
\textit{Remark 1\/} below),
\item $\Su\subset\underline{J_2}$,  projectable on
$\underline{J_1}$  ($\underline{J_1}\simeq\Su$), the embedding
in $\underline{J_2}$ of the presheaf of
$\theta_\Rset$-modules of germs of differential sections of
$J_2$, defined by the  system 
(\ref{third}) of algebraic equations at any given points
$x_0\in\Rset^n$ (not everywhere, as mentioned above),
\item $\tsms$, the presheaf of  
$\theta_\Rset$-modules of germs  of
global \mbox{1-forms} on $\Rset^n$.
\end{enumerate}
\end{definition}
Remark 1: Through this set
of definitions, we do not consider  PDEs solutions, but instead, solutions of
algebraic equations at any given point $x_0$. In this light, PDEs solutions are
to be regarded as particular ``coherent" subsheafs for which equations
(\ref{third}) are satisfied everywhere in $\Rset^n$, i.e., at $x\neq x_0$, and not
solely at  $x_0$. We insist that the algebraic equations
(\ref{third}) do not concern solutions of a PDE system, but the
values of  second derivatives of  functions at $x_0$, depending on
those of first order at most at $x_0$, with no  constraints between
first and zero-th order values of these functions at $x_0$.
\par\medskip
Then,  considering the local diffeomorphisms
\[(\wedge^k\,T^\ast\Rset^n\otimes_\Rset J_r)_{x_0}
\simeq(\{x_0\}\otimes_\Rset
J_r)\times(\wedge^k\,T^\ast_{\!x_0}\Rset^n
\otimes_\Rset J_r)\] 
with $0\leq k\leq n$ and $r\geq0$,
we set the definitions:
\begin{definition}We define the local operators:\par
\begin{enumerate}
\item
$j_1:(x_0,\rho_0)\in\Sz\longrightarrow
(x_0,\rho_0,\boldsymbol\rho_1,\boldsymbol{\rho}_2)\in\Su$
with 
$\boldsymbol\rho_1=
(\partial_1\rho_0,
\cdots,
\partial_n\rho_0)$ and
$\boldsymbol\rho_2=
(\partial_{11}^2\rho_0,
\partial_{12}^2\rho_0,
\cdots,
\partial_{nn}^2\rho_0)$,
\item 
$D_{1,c}:\boldsymbol{\rho}^2_0\equiv(x_0,\rho_0,
\boldsymbol{\rho}_1,\boldsymbol{\rho}_2)\in\Su
\longrightarrow(\boldsymbol{\rho}^1_0,\sg_0,
\sg_1,\cdots,\sg_n)
\in\tsms\otimes_{\theta_\Rset}
\underline{J_1}$,
with $\A$, $\B$ and $\boldsymbol{\rho}^1_0\equiv(x_0,\rho_0,
\boldsymbol{\rho}_1)$ satisfying
relations (\ref{rhoAB}), and $P_c=\{\sg_0,
\sg_1,\cdots,\sg_n\}$ being a Pfaffian system of linearly independent
regular 1-forms on $J_1$,
\item 
\(D_{2,c}:(\boldsymbol{\rho}^1_0,\sg_0,\sg_1,\cdots,\sg_n)
\in\tsms\otimes_{\theta_\Rset}
\underline{J_1}\longrightarrow
(\boldsymbol{\rho}^1_0,\zeta_0,\zeta_1,\cdots,\zeta_n)\in\)
\(\wedge^2\tsms\otimes_{\theta_\Rset}\underline{J_1}\), with
\begin{eqnarray*}
\zeta_0=\sum_{i,j=1}^n\I_{i,j}\,dx_0^i\wedge\,dx_0^j\,,\qquad
\zeta_k=\sum_{i,j=1}^n\J_{j,i,k}\,dx_0^i\wedge\,dx_0^j\,,
\end{eqnarray*}
the functions
$(x_0,\rho_0,\boldsymbol{\rho}_1,\boldsymbol{\rho}_2)\in\Su$
and the tensors
$\I$, $\J$, $\A$ and $\B$  satisfying the relations (\ref{ABIJeq}).
\end{enumerate}
\end{definition}
Then from all that preceeds, we can deduce:
\begin{theorem}
The differential sequence
\[
\begin{CD}
0@>>>\Sz@>j_2>>\Su
@>D_{1,c}>>\tsms\otimes_{\theta_\Rset}
\underline{J_1}@>D_{2,c}>>
\wedge^2\tsms\otimes_{\theta_\Rset}\underline{J_1}\,,
\end{CD}
\]
with the $\Rset$-linear local
differential operators $D_{1,c}$ and $D_{2,c}$\,,
is  exact (where the first injectivity, namely $j_1$, results from remark 1).
\end{theorem}
Remark 2: Before proceeding with the proof of this Theorem, a
few comments are in order.
{\rm
The continuation ``on the right" of the differential sequence  above
would require, in order to demonstrate the exactness, a
generalization of the Fr\"obenius Therorem to
$p$-forms with $p\geq2$, which, to our knowledge at least, is not
available in full generality, no more as the concept of canonical
contact $p$-forms representations. Indeed, the higher local
differential operators $D_{i\ge3,c}$ would be non-linear, in
contrary to the usual Spencer differential operators, because
of the non-linearity of the GHSS system. We are faced to the
same situation encountered in the Spencer sequences for Lie
equations, these sequences being truncated at this same order
two.} {\rm The sequence above is a physical gauge sequence, for
which we can make the following identification:
$\tsms\otimes_{\theta_\Rset}
\underline{J_1}$ is the space of the gauge \textit{potentials\/} $\A$ and $\B$, whereas
$\wedge^2\tsms\otimes_{\theta_\Rset}\underline{J_1}$ is the space of the gauge \textit{
strength fields\/}
$\I$ and $\J$.}
\par\medskip
This sequence is close to a kind of Spencer
linear sequence \cite{spencerlin69}. It differs essentially
in the tensorial product which is taken on  $\theta_\Rset$
(because of the non-linearity of the ``GHSS system", inducing
a $\rho$ ``dependence" of the various Pfaff forms) rather
than on the $\Rset$ field as is in the original linear Spencer
theory \cite{spencerlin69} (other developements have
included the $\theta_\Rset$ case after this first
Spencer original version). Also, since the system
$P_c$ is integrable, it is always, at least locally, diffeomorphic  to an
integrable set of Cartan 1-forms in
$T^\ast\Rset^n\!\otimes_\Rset\!{J_1}$
associated to a particular finite Lie algebra $g_c$ 
(of dimension greater or equal to
$n+1$), with corresponding Lie group
$G_c$ acting on the left on each leaf of the foliation $\mathcal{F}_1$
\cite{almeida2003,AT92,tT90}.  It follows  that the integrals in (\ref{deformpar})
would define  a deformation class in the first non-linear 
Spencer cohomology space of deformations of global sections from
$\Rset^n$ to a sheaf of Lie groups  $G_c$ \cite{kumperaspencer}
(see also \cite{lislereid}, though within a different approach).\par In
addition, 1) in Theorem 1, the fonction $u$ is defined with integrals
associated to the definition of a homotopy operator of the differential
sequence above \cite{buttin67}, and 2) in  Theorem 2, the metric
$\omega$ is allowed to be of class $C^\infty$, rather than analytic, as in 
Theorem 1, because  formal properties only are
considered.\par\medskip  
\textbf{Proof of Theorem 2:} At $\Sub_c^1$ the sequence
exactness is trivial and we may pass to the exactness of the
differential sequence at
$\tsms\otimes_{\theta_\Rset}\underline{J_1}$.\par
In a neighbourhood of an open set $V(\bfcM)\subset J_1$ of
$\bfcM\in J_1$, the condition $D_{2,c}(\sg)=0$ implies the
relations:
\begin{eqnarray}
&d\sgt_0=\sum_{i=1}^n\,dx_0^i\wedge\sgt_i\,,\label{dsg0}\\
&d\sgt_i=\sum_{j=1}^n\,dx_0^j\wedge\sgt_{ij}\label{dsgi}
\end{eqnarray}
with:
\begin{equation}
\sgt_{ij}=c_i\sgt_j+c_j\sgt_i-\omega_{ij}\Big\{k_0 e^{2c_0}\sgt_0+
\sum_{k,h=1}^n\omega^{kh}c_h\sgt_k\Big\}
+\sum_{k=1}^{n}\gamma^k_{ij}\sgt_k\,,
\label{fhs}
\end{equation}
the ``$\tilde\sg$" 1-forms  are defined above $J_1$, and
correspond to the 1-forms $\sg$ defined in a  neighbourhood
$W(X_0)\subset\Rset^n$ at $x_0\in W(X_0)$: they are such that if
$p_1: J_1\longrightarrow\Rset^n$ stands for the standard projection,
then $p_1(V(\bfcM))=W(X_0)$, $p_1(\bfcm)=x_0$ and
$p_1(\bfcM)=X_0$.\par Regularity and
linear independence, ensure the existence of a locally integrable
manifold $\mathcal{V}_1$, with dimension $n$, and of 
$n+1$ first integrals $\{y_\nu\}$
($\nu=0,1,\dots,n$). Up to  constants, the functions $y_\nu$
can be choosen such that
$y_\nu(\bfcM)=0$. Then, at
$\bfcM$, we have  the relations:
\begin{equation}
\tilde{\sg}_\nu(\bfcM)\equiv dy_\nu\!\big/_{\!\!\bfcM}\,,
\end{equation}
and in $V(\bfcM)$, the relations 
\begin{equation}
\tilde{\sg}_\nu=dy_\nu-\sum_{\mu\neq\nu}^{n} f_\nu^\mu(y)dy_\mu\,,
\label{desy}
\end{equation}
with $f_\nu^\mu(y)\longrightarrow0$ when $y\longrightarrow0$, that
is when $\bfcm\longrightarrow\bfcM$.\par These relations can also
be defined on the presheaves of the $J_1$ local
sections. This is because it exists a $C^1$-mapping, say $s$,
from  $W(X_0)$ into $V(\bfcM)$, such that
$s(W(X_0))=U(\bfcM)\subset V(\bfcM)$ and
$s(x_0)=\bfcm$. And thus locally, one has
$\mathcal{V}_1\cap U(\bfcM)\simeq W(X_0)$. In the relations
(\ref{desy}), it is therefore possible to take
$y_j\equiv c_j$ ($j=1,\dots,n$). Setting
$s^\ast(dy_j)=s^\ast(dc_j)\equiv dx^j_0$, and denoting by ``$\rho$"
the functions $\rho_j(x_0)=c_j$ and $\rho_0(x_0)=y_0$, associated
to $s$, we have immediately in particular
$\sg_0=s^\ast(\sgt_0)$, and
$\forall\,x_0$:
\begin{equation}
\sg_0\equiv d\rho_0-
\sum_{i=0}^{n}f_0^i(\rho)\,dx_0^i\,.
\label{sg0rho}
\end{equation}
We set
$\rho_{i}\equiv{}f_0^i(\rho)$.
Now, from (\ref{sg0rho}) and the pull-back of (\ref{dsg0}), one 
deduces that
\begin{equation}
\sum_{i=1}^{n}dx_0^i\wedge (d\rho_{i}-\sg_i)=0\,,
\label{exa}
\end{equation}
and in particular:
\begin{equation}
dx_0^1\wedge dx_0^2\wedge\dots\wedge dx_0^n\wedge
(d\rho_{i}-\sg_i)=0\,.
\end{equation}
Consequently
\begin{equation}
d\rho_{i}-\sg_i=\sum_{j=1}^{n}\rho_{i,j}\,dx_0^j
\Longleftrightarrow
\sg_i=d\rho_{i}-\sum_{j=1}^{n}\rho_{i,j}\,dx_0^j\,,
\end{equation}
that are alternatives to the pull-backs by $s$ of relations
(\ref{desy}). We thus have,
\begin{eqnarray}
&&\sg_0=
d\rho_0-\sum_{i=1}^{n}\rho_i\,dx_0^i\,,\label{sg0dr}\\
&&\sg_i=d\rho_i-\sum_{j=1}^{n}\rho_{i,j}\,dx_0^j\,.\label{sgidr}
\end{eqnarray}
Now, out of (\ref{sgidr}) and the pull-backs of
(\ref{dsgi}), one deduces also the relations,
\begin{equation}
\sum_{j=1}^{n}\,dx_0^j\wedge
d\rho_{i,j}=\sum_{j=1}^n\,dx_0^j\wedge\sg_{ij}\Longleftrightarrow
\sum_{j=1}^{n}\,dx_0^j\wedge
(d\rho_{i,j}-\sg_{ij})=0\,.
\end{equation}
By the same procedure as above, we thus get:
\begin{equation}
d\rho_{i,j}-\sg_{ij}=\sum_{j=1}^{n}\rho_{i,j,k}\,dx_0^k
\Longleftrightarrow
\sg_{ij}=d\rho_{i,j}-\sum_{k=1}^{n}\rho_{i,j,k}\,dx_0^k\,.
\end{equation}
Moreover, in view of (\ref{exa}) and (\ref{sgidr}) we deduce the
symmetries:
$\rho_{i,j}=\rho_{j,i}\equiv\rho_{ij}$ and
$\rho_{i,j,k}=\rho_{j,i,k}\equiv\rho_{ij,k}$. Then,
considering the coefficients of a same basis differential
form with $\boldsymbol{\rho}_2\equiv(\rho_{ij})$,
$\boldsymbol{\rho}_1\equiv(\rho_{i})$, and the system of
algebraic equations for  $\boldsymbol{\rho}^2_0$ deduced from
(\ref{dsg0}) and (\ref{dsgi}), we conclude that
$\boldsymbol{\rho}^2_0\in\Su$.
\hfill
$\Box$
\par\bigskip
In order to know the effects on
$\m$ of these infinitesimal deformations, we need to describe
what  are their incidences  upon the  objects acting
primarily on $\Rset^n$, namely the applications $\hf$. Thus,
we pass to the study of what we call the ``ab system" of the PDE system
(\ref{confeq}).
\par\medskip

%%%%%%%%%%%%%%%%%%%%%%%%%%%%%%%%%%%%%%%%%%%%%%%%%%%%%%
%%%%%%%%%%%%%%%%%%%%%%%%%%%%%%%%%%%%%%%%%%%%%%%%%%%%%%

\subsection{The ``ab system"} This system is defined
by the first two sets of PDE (\ref{f1}) and (\ref{f2}). 
For this system of Lie equations, we will begin with recalling well-known
results, but in the framework of the present context. Applying the same
reasoning than in the previous subsection, and considering the series (\ref{serief}) for
the
$\hf$,  we first obtain the following results, which hold up to order two: 
\begin{subequations}
\begin{eqnarray}
&&\sum_{r,s=1}^n\omega_{rs}(\hat{a}_0)
\,\,\hat{a}^{r}_{i}\,\hat{a}^{s}_{j}=
e^{2c_0}\omega_{ij}(x_0)\,,
\label{a1h}\\ 
&&\hat{a}^k_{ij}+
\sum_{r,s=1}^n{}\gamma^{k}_{rs}(
\hat{a}_0)\,\,\hat{a}^{r}_{i}\,\hat{a}^{s}_{j}=
\nonumber\\
&&\hspace{3cm}\sum_{q=1}^n{}\hat{a}_q^k\left(
\gamma^{q}_{ij}(x_0)+c_i\delta^q_j+
c_j\delta^q_i-\omega_{ij}(x_0)c^q\right)\,,
\label{a2h}
\end{eqnarray}
\label{a12h}
\end{subequations}\par\noindent
which clearly show that, fixing the $c$'s, $J_1(\Rset^n)$ is
diffeomorphic to an embedded submanifold of the 2-jets
affine bundle $J_2(\Rset^n)$ of the $C^\infty(\Rset^n,\Rset^n)$
differentiable applications on
$\Rset^n$. In second place, we get relations, from the (T) 
system (cf. section 2), for
the coefficients of order 3 that we write as
($\hat{a}_1\equiv(\hat{a}^i_j)$;
$\hat{a}_2\equiv(\hat{a}^i_{jk})$,\dots,
$\hat{a}_k\equiv(\hat{a}^i_{j_1\cdots j_k})$;
$\bfah^k_0\equiv(\hat{a}_0,\cdots,\hat{a}_k)$):
\begin{equation}
\hat{a}^i_{jkh}\equiv
\hat{A}^i_{jkh}(x_0,\bfah^2_0)\,,
\label{a3h}
\end{equation}
where $\hat{A}^i_{jkh}$ are algebraic functions,
pointing out in this expression the independence on
the ``$c$" coefficients (as in the relations (\ref{a1h}), for
instance, when  $c_0$ is expressed in terms of the determinant of
$\hat{a}_1$). We denote by
$\Oc^i_J$ the Pfaff 1-forms at $x_0$ and $\{\hat{a}\}$
(or at  $(x_0,\{\hat{a}\})$):
\begin{equation}
\Oc^i_J\equiv
d\hat{a}^i_J-\sum_{k=1}^n\hat{a}^i_{J+1_k}dx^k_0\,,
\label{Oma}
\end{equation}
and  setting the $\hat{a}$'s as values of
functions $\hat{\tau}$ depending on $x_0$ (in some way we
make a pull-back on $\Rset^n$), we define the tensors $\kc$ by:
\begin{equation}
\Oc^i_J\equiv\sum_{k=1}^n\Big(\partial^0_k\hat{\tau}^i_J-
\hat{\tau}^i_{J+1_k}\Big)\,dx^k_0\equiv
\sum_{k=1}^n\kc^i_{k,J}\,
dx^k_0\,.
\label{omtau}
\end{equation}
Then from the relations:
\[
\e^{2c_0}\oms{r}{s}(\hat{a}_0)=
\sum_{i,j=1}^n\oms{i}{j}(x_0)\hat{a}^r_i
\hat{a}^s_j\,,\quad
\sum_{i=1}^n\gam{i}{i}{k}=
\frac{1}{2}\sum_{i,j=1}^n\oms{i}{j}\,\partial^0_k\omi{i}{j}\,,
\]
we deduce from (\ref{a1h}) with $\hat{b}\equiv\hat{a}_1^{-1}$, that the
$\Oc^i_j$ 1-forms satisfy, at $(x_0,\bfah^1_0)$, the relations:
\begin{equation}
\widehat{H}_0(x_0,\bfah^1_0,\Oc^k_L;|L|\leq1)
\equiv\sum_{i,j=1}^n\hat{b}^j_i\,\Oc^i_j+
\sum_{j,k=1}^n\gamma^j_{jk}(\hat{a}_0)\Oc^k
=n\sg_0\,.\label{nsg0}
\end{equation}
Similar computations show that the 1-forms $\sg_i$
can be expressed as quite long relations, linear in the
$\Oc^j_J$ ($|J|\leq2$), with coefficients which are algebraic functions
depending on the $\hat{a}_K$ ($|K|\leq2$), the derivatives of the metric
and the Riemann-Christoffel symbols, all of them taken 
either at $x_0$ or $\hat{a}_0$. Then, we set:
\begin{equation}
\sg_i\equiv\widehat{H}_i(x_0,\bfah^2_0,\Oc^j_I;
|I|\leq2)\,.
\label{nsgi}
\end{equation}
From (\ref{a3h}) the 1-forms 
$\Oc^i_{jkh}$  are also sums of 1-forms
$\Oc^r_K$ ($|K|\leq2$) with the same kind of coefficients
and not depending on the $\sg$'s, and we write (without any
more details since  it is not necessary for our
demonstration below):
\begin{equation}
\Oc^i_{jkh}\equiv\widehat{K}^i_{jkh}(x_0,
\bfah^2_0,\Oc^r_K;|K|\leq2)\,,
\label{K3}
\end{equation}
where $\widehat{K}^i_{jkh}$ are functions which are linear in the 1-forms
$\Oc^r_K$.\par\medskip
Let us denote by $\widehat{\text{\eu P}}_2\subset J_2(\Rset^n)$ the set of
elements $(x_0,\bfah_0^2)$ satisfying the relations
(\ref{a12h}) whatever are the $c$'s.  Then the Pfaff system
we denote
$\widehat{P}_2$ over $\widehat{\text{\eu P}}_2$ and generated by
the 1-forms $\Oc^j_K\in{T^\ast\Rset^n}
\otimes_\Rset{J_2(\Rset^n)}$
in (\ref{Oma}) with $|K|\leq2$, is locally integrable on
every neighborhood
$U{(x_0,\bfah_0^2)}{\subset}J_2(\Rset^n)$, since at $(x_0,\bfah^2_0)$
we have ($|J|\leq2$):
\begin{equation}
d\Oc^i_J-\sum_{k=1}^n dx^k_0\wedge\Oc^i_{J+1_k}\equiv0\,,
\label{dom}
\end{equation}
together with (\ref{K3}).\par\medskip
Let us now consider the ``Poincar\'e system" whose 
corresponding notations will be free of ``hats". We denote by
$\Omega^i_J$ the Pfaff 1-forms corresponding to this
system, i.e. the system defined by the PDE
(\ref{f1}) and (\ref{f2}) with a vanishing function
$\alpha$. The corresponding 1-forms ``$\sg$"  are also
vanishing everywhere on $\Rset^n$ and the $\Omega^i_J$
satisfy all of the previous relations, but  with the $\sg$'s
cancelled out. Then it is easy to
see that the 
$\Omega^i_J$ 1-forms ($|J|\geq2$) are generated
by the set of 1-forms $\Omega^j_K$ ($|K|\leq1$);
We have in particular
\begin{equation}
\Omega^k_{ij}=-\left\{\sum_{r,s,h=1}^n(\partial^0_h
\gamma^k_{rs})({a}_0)\Omega^h\,{a}^r_i\,{a}^s_j
+\sum_{r,s=1}^n\gamma^k_{rs}({a}_0)[\,{a}^r_i\,\Omega^s_j+
{a}^s_j\,\Omega^r_i\,]\right\}\,,
\label{K2}
\end{equation}
with $(x_0,\bfah_0^1\equiv \bfa_0^1)\in\text{\eu P}_1\subset
J_1(\Rset^n)$,  $\text{\eu P}_1$ being the set of elements
satisfying the relations (\ref{a1h}) with $c_0=0$. Similarily
the Pfaff system we denote by ${P}_1$ over $\text{\eu P}_1$
and generated by the 1-forms
$\Omega^j_K$ in (\ref{Oma}), but free of hats and with $|K|\leq1$, is
locally integrable on every neighborhood
$U{(x_0,\bfa_0^1)}\subset\text{\eu P}_1$, since at the point
$(x_0,\bfa_0^1)$ we have  the relations
(\ref{dom}) with $|J|\leq1$ together with  relations
(\ref{K2}).\par
Then, considering  $J_1(\Rset^n)$ embedded in $J_2(\Rset^n)$, as
well as  $\text{\eu P}_1$ in $\widehat{\text{\eu P}}_2$, and
defining 
$\text{\eu P}_2\subset\widehat{\text{\eu P}}_2$ as the set of elements $(x_0,\bfa_0^2)$
satisfying the relations (\ref{a12h}) with $c_0=c_1=\dots=c_n=0$, we obtain the
following theorem justifying the structure's deformation point of
view:
\begin{theorem}
The sequence
\begin{equation}
\begin{CD}
0@>>>P_1@>b_1>>\widehat{P}_2@>e_1>>P_c@>>>0\,.\label{pseq}
\end{CD}
\end{equation}
is a local exact
splitted sequence over $\text{\eu P}_2$.
\end{theorem}
In this sequence a back-connection
$b_1$  and a connection
$c_1:P_c\longrightarrow\widehat{P}_2$ are such that
($|J|\leq2$):
\begin{equation}
\Oc^i_J=\Omega^i_J+\chi^i_J(x_0,\bfa^2_0)\,\sg_0+
\sum_{k=1}^n\chi^{i,k}_J(x_0,\bfa^2_0)\,\sg_k\equiv 
\Omega^i_J+c^i_1(\sigma_K;|K|\leq1)\,,\quad
\label{splitom}
\end{equation}
with $(\Omega^i_{jk})=b_1(\Omega_J;|J|\leq1)$ satisfying (\ref{K2}) for any given 
$\Omega^h_J$ with $|J|\leq1$, and where the tensors $\chi$ are defined on
$\text{\eu P}_2$. The maps $b_1$ and $c_1$  define  the projective map $e_1$ if
the tensors $\chi$  satisfy the relations:
\begin{subequations}
\begin{eqnarray}
&&\widehat{H}_0(x_0,\bfa^1_0,\chi^k_L;|L|\leq1)=n\,,\qquad
\widehat{H}_0(x_0,\bfa^1_0,\chi^{k,i}_L;|L|\leq1)=0\,,\\
%\hspace{-2cm}
&&\widehat{H}_i(x_0,\bfa^2_0,\chi^k_L;|L|\leq2)=0\,,1\qquad
\widehat{H}_i(x_0,\bfa^2_0,\chi^{k,h}_L;|L|\leq2)=
n\delta^h_i\,,
\end{eqnarray}
\label{backon}
\end{subequations}
in order to preserve the relations (\ref{nsg0}) and 
(\ref{nsgi}),  i.e. $e_1\circ c_1=id$.

%%%%%%%%%%%%%%%%%%%%%%%%%%%%%%%%%%%%%%%%%%%%%%%%%%%%%%
%%%%%%%%%%%%%%%%%%%%%%%%%%%%%%%%%%%%%%%%%%%%%%%%%%%%%%
%%%%%%%%%%%%%%%%%%%%%%%%%%%%%%%%%%%%%%%%%%%%%%%%%%%%%%

\section{The  spacetime $\m$ unfolded by 
Gravitation and Electromagnetism}
From now on, we
consider the relations (\ref{splitom}) with $|J|=0$ and
the $\Oc^i$ as fields of {\em``tetrads"\/}. Then we get a  metric
$\nu$ for the {\it``unfolded spacetime manifold $\m$"}
defined at $x_0$, and corresponding to the metric $g$ at $p_0$ in
$\m$:
\begin{eqnarray*}
&&\nu(x_0)=(\omega+\delta\omega)(x_0)=
\sum_{i,j=1}^n\omega_{ij}\circ\hat{\tau}(x_0)\,
\Oc^i(x_0)\otimes\Oc^j(x_0)\,,\\
&&\Oc^i(x_0)=d\hat{\tau}^i-\sum_{k=1}^n\hat{\tau}^i_k(x_0)\,dx_0^k\equiv\sum_{k=1}^n\kc^i_{k}(x_0)\,
dx^k_0\,,\\
&&\kc^i_k(x_0)=\kappa^i_k(x_0)+\chi^i(x_0,\bftau_0^2)\,\A_k(x_0)+
\sum_{h=1}^n\chi^{i,h}(x_0,\bftau_0^2)\,\B_{k,h}(x_0)\,.
\end{eqnarray*}
We consider the
particular case for which the S-admissible metric
$\omega$ is equal to\linebreak
$\text{diag}[+1,-1,\cdots,-1]$ (and thus
$k_0=0$), the $\chi$'s are independent on $x_0$ and the
$\tau$'s since in view of relations (\ref{backon}) the independence of the zero-th
order $\chi$'s can be consistently assumed, and
$\kappa^i_j=\delta^i_j$,  i.e., the deformation of
$\omega$ is only due to the tensors $\A$ and
$\B$. Thus, one has the general relation between
$\nu$ and $\omega$: $\nu=\omega\,+$ linear and quadratic
terms in $\A$ and $\B$, and from the metric
$\nu(x_0)$, one can deduce the Riemann and Weyl curvature
tensors of the {\it``unfolded spacetime $\m$"\/} at $x_0$.\par
Under these assumptions, we can also define the dual vector
fields $\widehat\partial$ such that at first order
$\Oc^i(\widehat\partial_j)\simeq\delta^i_j$. We have, the
relations:
\[
\widehat\partial_j=\sum_{q=1}^n\beta^q_j(x_0)\,\partial^0_q\,,
\]
with
\[
\beta^q_j(x_0)=\delta^q_j(x_0)-\chi^q(x_0,\bftau_0^2)\,\A_j(x_0)-
\sum_{k=1}^n\chi^{q,k}(x_0,\bftau_0^2)\,\B_{j,k}(x_0)\,.
\]
In view of making easier computations for a relativistic action
deduced from the metric tensor
$\nu$, we consider this metric in a {\it ``weak fields limit"\/},
where the metric $\nu$ is linear in the tensors $\A$  and $\B$,
and where the quadratic terms are neglected. It follows
that in a wide part of tensorial expressions, the
derivatives $\widehat\partial$ can be approximated by the 
$\partial^0$ ones.
Furthermore, from  relations (\ref{ABIJeq}) and  taking
into account the latter approximation, we have:
\[\partial^0_i\A_k-\partial^0_k\A_i\simeq\B_{k,i}-\B_{i,k}=
\F_{ik}\,,
\quad
\partial^0_j\B_{k,i}-\partial^0_k\B_{j,i}\simeq
0\,,\quad\J_{j,k,i}\simeq\partial^0_k\B_{j,i}\,,
\]
since  the functions $\rho$ take also small
values in this assumed weak fields limit. We can therefore write 
\(
\nu_{ij}\simeq\omega_{ij}+\epsilon_{ij}\,
\),
where the  coefficients $\epsilon_{ij}$ can be considered as small
perturbations of the metric field $\omega$. Now, a most important point
comes about when considering that this perturbation is (linearly) 
constructed out of 
$\A$, $\B$ and  $\chi$ tensors, realizing, in view of relations
(\ref{rhoAB}) and (\ref{firstset}), an explicit and non-trivial
unification of the electromagnetic and gravitational aspects valid for
all $n\geq4$.\par
It is worth noticing that this feature is conserved by the full
non-linearized expressions, though under a more complicated form, and
that a numerical resolution for the $\nu$ metric field is certainly
worth looking for.\par 
This unification is most definitely at variance with the ones we are
used to, based on superstring field theories, and in some sense just
goes the other way round. Still it opens over a wide range of quite
unexpected interpretations and/or speculations, some of them to be
quickly evoked shortly. We think that this formalism could also shed
an interesting new ligth on enigmas or difficulties such as those
listed in \cite{jmllsphinx}, for example.\par  Now, we can mention that from (\ref{LAB}), the
tensor $\A$  satisfies  the well-known system of PDE with the Lorentz gauge
condition ($\square^0$ being the d'Alembertian with respect to $x_0$):
\begin{equation}
\square^0\A_i=\widetilde{\mathcal{J}}_i\,,
\label{galo}
\end{equation}
where $\widetilde{\mathcal{J}}$ is an $n$-current. On the other hand, from
a Lagrangian density of type (\ref{LABIJ}) such as (with uppering
and lowering of indices operated by $\omega$ at first order):
\[
\Lag\equiv\sum_{j,k=1}^{n}\B_{j,k}\widetilde{\mathcal{K}}^{j,k}+
\frac{1}{2}\sum_{i,j,k=1}^{n}\J_{j,[k,i]}\J^{j,[k,i]}
\]
where
\[
\J_{j,[k,i]}=\J_{j,k,i}-\J_{j,i,k}\,,
\]
an analogous  PDE system results:
\begin{equation}
\square^0\B_{j,i}=\widetilde{\mathcal{K}}_{j,i}
\label{gegalo}
\end{equation}
with  ``gauge conditions":
\[
\sum_{i=1}^n\partial^0_i\B^{i,j}=0\,,
\]
and a ``generalized $n$-current" $\widetilde{\mathcal{K}}$. Now,
a most interesting aspect is that from the PDE systems (\ref{galo})
and (\ref{gegalo}), it is possible to calculate a metric field
$\nu$, and in the static case, a Newtonian potential, linearly
depending on $\A$ and/or $\B$, and satisfying   Poisson  equations.
This  means that in our  approach,   the Newtonian limit
is reached without any need for Einstein equations to be satisfied
by  the metric field $\nu$, the latter being replaced by the
Euler-Lagrange equations deduced from (\ref{LABIJ}) together with
the first set of differential equations (\ref{firstset}) !\par
Eventually, we will end up this section with a few speculations
concerning the point base $p_0$ motion in a spacetime endowed with
the metric field
$\nu$.
\par\medskip
Let
$i$ be a differential map
$i:s\in[0,\ell]\subset\Rset\longrightarrow
i(s)=x_0\in\Rset^n\simeq\m$, and 
$U(s)\equiv di(s)/ds\,$,
 such as $\nu(U,U)\equiv{\parallel}U{\parallel}^2=1$. We define the
relativistic action
$S_1$ by:
\[
S_1=\int_0^\ell\sqrt{\nu(U(s),U(s))}\;ds
\equiv\int_0^\ell \sqrt{L_\nu}\;ds\,.
\]
We also take the tensors
$\chi$ as depending on $s$ only. The Euler-Lagrange equations
for the Lagrangian density $\sqrt{L_\nu}$ are not
independent because $\sqrt{L_\nu}$ is a homogeneous
function of degree 1, and thus satisfies an additional
homogeneous differential equation. Then, it is well-known
that the variational problem for
$S_1$ is equivalent to consider the variation of the  
action $S_2$ defined by
\[
S_2=\int_0^\ell{\nu(U(s),U(s))}\;ds
\equiv\int_0^\ell L_\nu\;ds\,,
\]
but constrained by the condition $L_\nu=1$. In this
case,  this shows that $L_\nu$ must be
considered, firstly,  with an associated Lagrange multiplier, namely a
mass, and secondly, that the $L_\nu$ explicit expression with respect
to $U$ will appear only in the variational calculus. In the
weak fields limit, we obtain:
\begin{eqnarray}
\qquad L_\nu&=&\omega(U,U)+
2\sum_{j,k=1}^n\omega_{kj}\,\chi^k\,U^j\;.\;
\sum_{i=1}^n\A_i\,U^i\nonumber\\
&&\hspace{3cm}+2\sum_{j,k,h=1}^n\chi^{k,h}\,\omega_{kj}\,U^j\;.\;
\sum_{i=1}^nU^i\,\B_{i,h}\,.
\label{lnu}
\end{eqnarray}
From the latter relation, we can deduce a few
physical consequences among others. On the one hand, 
if we denote
by ($h=1,\cdots,n$)
\begin{subequations}
\begin{eqnarray}
&&\Cc^h(\chi,U)\stackrel{\,\,def.}{=}
\sum_{j,k=1}^n\omega_{kj}\chi^{k,h}U^j\equiv\zeta^h\,,
\label{thomgen}\\
&&\Cc^0(\chi,U)\stackrel{\,\,def.}{=}
\sum_{k,j=1}^n\omega_{kj}\chi^kU^j\equiv\zeta^0\,,
\label{thomas}
\end{eqnarray}
\label{prec}
\end{subequations}\par\noindent
then we recover in
(\ref{lnu}), up to some suitable constants, the
Lagrangian density for a particle (with charge
$\zeta^0$), with the {\it velocity n-vector $U$\/}
(\({\parallel}U{\parallel}^2=1\)), embedded in an
external electromagnetic field. But also from the relation
(\ref{thomas}) we will find {\it``a generalized Thomas
precession"\/} if  the tensor $(\chi^k)$, assumed 
to depend on $s$, in that specific
case only, but not on $x_0$,  is ascribed (up to a suitable constant
for units) to a {\it``polarization
$n$-vector"\/} \cite[p.~270]{bacry} {\it``dressing"\/} the
particle (a spin for instance). Likewise,  the tensor
$(\chi^{k,h})$ might be a \textit{polarization tensor\/} of some
matter,  and again, the particle would be ``dressed" with this kind
of polarization.\par\medskip
More generally, the Euler-Lagrange equations associated to
$S_2$ would define a system of ``local" geodesic equations 
with Riemann-Christoffel symbols $\Gamma$ and such
that (with \(\nu^{ij}\simeq\omega^{ij}\) at first order 
and recalling that the $\chi$'s are
constants)
\begin{equation}
\frac{\;\,dU^r}{ds}=-\sum^n_{j,k=1}
\Gamma_{jk}^r\;U^j\,U^k
+\zeta^0\sum_{i,k=1}^n\omega^{kr}\,{\F_{ki}}\,U^i\,,
\label{geod}
\end{equation}
with
\[
\Gamma_{jk}^r(x_0)=\frac{1}{2}\,
\left\{
\chi^r\,\left(\partial^0_k\A_j+\partial^0_j\A_k\right)+
\sum_{\ell=1}^n\chi^{r,\ell}
\left(\partial^0_j\B_{k,\ell}+\partial^0_k\B_{j,\ell}\right)
\right\}
\,.
\]
Let us indicate that we can compare (\ref{geod}) with the analogous
equation ($6.9''$) in \cite{fulton62} but with different
Riemann-Christoffel symbols.\par Moreover $\A$ and $\B$ must
satisfy \textit{the first\/} and \textit{second sets of
differential equations associated to\/} $\m$, i.e., the relations
(\ref{firstset}) and (\ref{FG}) but into which the derivatives
$\partial^0_j$ are substituted by the derivatives
$\widehat\partial_j$. Nevertheless in the weak fields limit these
equations reduce again to the equations (\ref{firstset}) and
(\ref{FG}).\par\medskip

The tensor $\Gamma$ would be associated to gravitational fields,
also providing other physical interpretations for the tensors
$\chi$. This tensor can also satisfy the
so-called ``meshing assumption" of Ghins and Budden \cite{ghins}. In
the present context, for $x_0$ restricted to  a
free-falling worldline $\mathcal{W}$, it involves in full generality
that we must have the relation:
$\Gamma^i_{jk}(x_0)=\gamma^i_{jk}(x_0)$. In particular, with
the choice taken for $\omega$ we have
$\Gamma^i_{jk}(x_0)=0$. Then, as required by the latter assumption and quoting
Ghins and Budden,  we would deduce on $\mathcal{W}$ only  that, indeed,
\textit{``special relativistic laws \emph{[would]} hold in
\emph{[their]} standard vectorial forms"\/} with relations
(\ref{firstset}), (\ref{FG}) and equation (\ref{geod}), for an
interacting particle in a non-necessarily locally flat spacetime. This meshing
assumption is fundamental since it prevents one from considering
the metric $\nu$ as a pull-back of $\omega$ in anyway, but rather
like a deformation, as we did in the present paper. Indeed in this
pull-back case,  at any point
$x_0$, the meshing condition would be
always satisfied from the definition of a metric applied on vectors
at $x_0$ (and would no  longer be an assumption !), i.e., we would
have everywhere only special relativity laws, even in a non-flat
spacetime (see the deep Ghins and Budden paper).
\par\medskip Coming back to  equations (\ref{geod}), the latter are
deduced irrespective of the conditions
(\ref{prec}) to be set to constants. Now, if the $\zeta$'s
are constants and in the case  of an explicit
$s$ dependence of the $\chi$'s, not the one induced by  $x_0=i(s)$,
this would lead to a modification of the action
$S_2$ resulting from the introduction of  Lagrange
multipliers $\lambda_0$ and $\lambda_k$ ($k=1,\cdots,n$)
in the Lagrangian density definition. We would then
define a new action of the type:
\[
{S}_2=\int_0^{\ell}
\left\{m{\parallel}U{\parallel}^2+
\sum_{i=1}^n\epsilon_{ij}\,U^j\,U^i
-\sum_{k=0}^n\lambda_k\,\Cc^k(\chi,U)
\right\}ds\,.
\]
The associated Euler-Lagrange equations
would be analogous to (\ref{geod}), but with additional terms coming
from the generalized Thomas precession  previously evoked.
Moreover, since we have the constraint
${\parallel}U{\parallel}^2=1$, we need a new Lagrange
multiplier denoted by $m$.\par
Then the variational calculus
would also lead to additional precession equations giving rise to
torsion. In the present situation, torsion is
not related to  unification but  to parallel
transports on manifolds which is a well-known geometrical
fact \cite{dieudon}. Hence, the existence of a
precession phenomenon for a spin or polarization $n$-vector
$(\chi^k)$ would be correlated with the existence of linear ODE for
a charged particle of charge $\zeta^0$ interacting with an
electromagnetic field. Otherwise, without (\ref{thomas}) the
ODE's would be non-linear and there wouldn't be  any kind of
precession  of any spin or polarization
\mbox{$n$-vector}.\par Consequently the  motion defined by
the second term in the r.h.s. of (\ref{geod}) for a spinning
charged particle  would just be, in this model,  a point of
view  resulting from an implicit separation of rotational and
translational degrees of freedom achieved by the specialized
(sensitive to particular subgroups of the  symmetry group of
motions) experimental apparatus in
$T_{p_0}\m$. This separation would insure either some simplicity
(i.e., linearity) or, since the measurements are achieved in
$T_{p_0}\m$, that the equations of motion are associated (\textit{via\/}
some kind of projections inherent to implicit dynamical
constraints, due to the experimental measurement process and
apparatus, fixing,  for instance,
$\zeta^0$ to a constant) to linear representations of 
tangent actions of the Lorentz Lie group on $T_{p_0}\m$. In the latter
case, one could say, a somewhat provocative way of course, that
special relativity invariance would have to be satisfied,
\textit{as much as possible\/}, by physical laws. Note that this ``reduction" to
linearity can't be done on the first summation in the r.h.s. of
(\ref{geod}), which must be left quadratic contrarily to the second
one, since the Riemann-Christoffel symbols can't be defined
covariantly (other arguments can be found
in~\cite{fulton62}).\par To conclude, equations (\ref{geod})
would provide us with another interpretation of spin or
polarization (the
$\chi$'s) as an object allowing moving particles to generate
effective spacetime deformations, as ``wakes" for instance.

%%%%%%%%%%%%%%%%%%%%%%%%%%%%%%%%%%%%%%%%%%%%%%%%%%%%%%
%%%%%%%%%%%%%%%%%%%%%%%%%%%%%%%%%%%%%%%%%%%%%%%%%%%%%%
%%%%%%%%%%%%%%%%%%%%%%%%%%%%%%%%%%%%%%%%%%%%%%%%%%%%%%

\section{Conclusion}
In the present article, we have been using the Pfaff sytems theory and the Spencer
theory of differential equations, to study the formal solutions of the
conformal Lie system with respect to the Poincar\'e one. More precisely,
we determined the difference between these two sets of formal solutions.
We gave a description of a ``relative" set of PDE, namely the ``GHSS
system", which provides the basis of a deployment from the Poincar{\'e} Lie pseudogroup
to a  sub-pseudogroup of the conformal Lie pseudogroup.  We studied
these two systems of Lie equations because of their specific occurrence in
physics,  particularly in electromagnetism as well as in Einsteinian
relativity.\par Relying on this concept of deployment, we made the
assumption that the unfolding is related to the existence of two kinds of
spacetimes, namely, a  substratum spacetime $\Sub$, from
which the  spacetime manifold $\m$ is unfolded. We recall that
not all of the given metrics on $\Sub$ are admissible so as to define, at least along
the lines proposed in the present article, such a deployment of $\m$ out of a
substratum spacetime $\Sub$. In the case of a substratum spacetime
$\Sub$ endowed with an appropriate S-admissible metric, allowing for unfolding,
we assumed that $\Sub$ is  equivariant with respect to  the conformal 
and  Poincar\'e pseudogroups, and  set its Riemannian scalar curvature 
to  a constant $n(n-1)k_0$, and its Weyl tensor to zero. At this stage, the
deployment evolution can be trivial or not depending on the  occurences of
spacetime singularities (of the deformation potentials)
parametrizing or dating what can be considered somehow as a kind of  spacetime
history. The deformation  potentials are built out of  a particular
relative Spencer  differential sequence associated to the ``GHSS system", and
describing smooth deformations of $\Sub$. Then a ``local" metric
$\nu$, defined on a moving tangent spacetime $\tom$ to the unfolded
spacetime $\m$, is constructed out of the S-admissible substratum
metric $\omega$, and of the deformation potentials. An unification
of two of the most fundamental aspects of our physical word come
out realized with, we think, a number of new and interesting new
lights shed on various issues of contemporary physics.\par The
tangent spacetimes dynamics  are given by a system of PDE satisfied
by their  Lorentzian  velocities
$n$-vectors $U$,  exhibiting both classical electrodynamic and ``local" (since metrics
are local) geodesic navigation in a spacetime endowed with gravitation.\par\medskip
Throughout, our approach has remained  ``classical'' and
quantization doesn't seem to play much role. But quite on the
contrary, we think that the formalism developed here could provide
the basis of a new and deeper recasting of whole branches of
physics, the quantic world included.  This is of course because of
the fundamental role played by symmetries at any scale, and not
solely at the classical one. In this line of thinking,  it is worth
pointing out some recent reflexions of G. 't Hooft about
``Obstacles on the Way Towards Quantization of Space, Time and
Matter"
\cite{thooftphil}. 't Hooft's  theory of ``Ontological States", involved in his
approach of ``deterministic quantization", strongly requires such a description of
spacetime as a fluid, as well as its close relationship with the principle of
coordinate invariance.

%%%%%%%%%%%%%%%%%%%%%%%%%%%%%%%%%%%%%%%%%%%%%%%%%%%%%%
%%%%%%%%%%%%%%%%%%%%%%%%%%%%%%%%%%%%%%%%%%%%%%%%%%%%%%
%%%%%%%%%%%%%%%%%%%%%%%%%%%%%%%%%%%%%%%%%%%%%%%%%%%%%%

\appendix
\section{Appendix}
It is interesting to remark that the Taylor coefficients could
be defined, for any solution, from partial as well as
covariant derivatives of the solution; The results remain essentially the
same. Indeed, given a covariant derivative
$\widetilde\nabla$ on
$\tom$, the basis vectors $e_i$ associated to the
coordinates $\xi^i=x^i-x^i_0$, the notations
$\widetilde\nabla^j\equiv(\widetilde\nabla)^j$ and
$\partial^j\equiv(\partial)^j$ for derivatives of order $j$, and the
monomial  $m\equiv
k\,(x^1-x_0^1)^{i_1}\dots(x^n-x_0^n)^{i_n}$ (where $k$ is an
element of an $\Rset$-vector space of finite rank type), the following 
relation at $x_0$ holds true:
\[
\frac{1}{i_1!\dots i_n!}\,\widetilde\nabla_{e_1}^{i_1}
\dots\widetilde\nabla_{e_n}^{i_n}(m)=\frac{1}{i_1!\dots
i_n!}\,
\partial_{1}^{i_1}\dots\partial_{n}^{i_n}(m)=k\,.
\]
This result is nothing but an example of a more general situation
such as the one encountered in the Kumpera-Spencer property
\cite[p. 70]{kumperaspencer} \cite[p. 34]{kumpspenCRM} or in the
Gasqui Lemma
\cite[Lemma 0.1]{gasqui} \cite[Lemma 0.2]{gasquigold}. This
shows also the vectorial (not affine) feature of the
symbol spaces, i.e., the space of elements $k$ at  any
given order
$|I|=i_1+\dots+i_n$. \par This property has also much to do with the so-called
``Meshing Assumption" of Ghins and Budden exhibited here as a
ground mathematical property, necessarily satisfied in the
framework of our gravity approach.\par Moreover,
in full generality we will not put any restrictions on the
kind of vectors $\xi$, the former to be defined by some
constraints involving covariant derivatives for instance. In
that case, we would consider for example, other Taylor
coefficients by substituting the
$c_{ij}-\sum_{\ell=1}^n\gamma^{\ell}_{ij}\,c_{\ell}$\, for
the $c_{ij}$. And then, considering $\alpha_i$  to be invariant
along a geodesic curve associated to the
basis vector $e_i$, we would deduce
$(c_{ij}-\sum_{\ell=1}^n\gamma^{\ell}_{ij}\,c_{\ell})\,\xi^i=0$.
This would simplify the Taylor expansion. Nevertheless, the
discussion in what follows would be  complexified, setting a
supplementary restrictive assumption on
$\alpha$ (for a detailed discussion on that point,
one may see \cite{rindpenbook1}).\par\bigskip\bigskip
%%%%%%%%%%%%%%%%%%%%%%%%%%%%%%%%%%%%%%%%%%%%%%%%%%%%%%
%%%%%%%%%%%%%%%%%%%%%%%%%%%%%%%%%%%%%%%%%%%%%%%%%%%%%%
%%%%%%%%%%%%%%%%%%%%%%%%%%%%%%%%%%%%%%%%%%%%%%%%%%%%%%
\bibliographystyle{plain}
%\bibliography{FD_Ocean_Biblio}

\end{document}